uuencoded %
file%

\documentstyle[preprint,eqsecnum,aps,epsf]{revtex}
\input epsf
\begin{document}
\draft
\title{Background thermal depolarization of electrons in storage
rings}
\author{A.C.C.\ Guimar\~aes~\footnotemark
\footnotetext{e-mail: acandido@ift.unesp.br}, G.E.A.\ Matsas~\footnotemark
\footnotetext{e-mail: matsas@axp.ift.unesp.br} 
and D.A.T.\ Vanzella~\footnotemark
\footnotetext{e-mail: vanzella@axp.ift.unesp.br}}
\address{
Instituto de F\'\i sica Te\'orica, Universidade Estadual Paulista\\
R. Pamplona 145, 01405-900 - S\~ao Paulo, S\~ao Paulo\\
Brazil}
\maketitle
\begin{abstract}
We discuss the influence of the background thermal bath on the
depolarization of electrons in high-energy storage rings, and on
the photon emission associated with the spin flip.  We focus, in
particular, on electrons at LEP. We show that in a certain interval
of solid angles the photon emission is
enhanced several orders of magnitude because of the presence of
the thermal bath. Notwithstanding,
the overall depolarization induced by the background 
thermal bath at LEP conditions is much smaller than the one 
induced  by plain acceleration  at zero-temperature and can 
be neglected in practical situations. Eventually we discuss 
in what conditions the background thermal bath can enhance the
overall depolarization by several orders of magnitude.
\end{abstract}
\pacs{41.60.-m, 12.20.Ds, 12.20.Fv}
\narrowtext
\section{Introduction}
\label{sec:intro}

Evidences of polarization in a single
circulating beam were detected unambiguously in the early 70's 
from Novosibirsk and Orsay \cite{B}.  Later, it was
observed in the storage ring SPEAR at Stanford  
a polarization  of $P\approx 76 \%$ 
\cite{LRS} and more recently a polarization of  $P\approx 90 \%$ 
\cite{Jetal}. The first observation of transverse beam polarization
in LEP was in 1990 \cite{Ketal}, reaching further $P\approx 57 \%$
\cite{Aetal}. 
Transverse and longitudinal polarization signals are
being observed since then (see e.g. \cite{Betal} and references
therein), and their utilization to test possible extensions to
the standard model constitutes source of excitement (see e.g.
\cite{CTRS}).  In spite of the peculiarities of the
different machines, theoretical
calculations indicate that the maximum natural transverse 
polarization possible to be reached by
ultra-relativistic electrons moving circularly 
in storage rings {\em at zero-temperature} is
$P\approx 92 \%$ \cite{ST}--\cite{J}. The main reason why the
polarization obtained is not complete is
the high acceleration under which these electrons 
are subjected. However, there are other sources of depolarization
which should be taken into account (see e.g. \cite{Tre}).

Here we discuss the contribution of the {\em background thermal bath}
on the depolarization of high-energy electron beams at storage
rings, and on the photon emission associated with the spin flip.
We focus on electrons at LEP, but our conclusions
will remain basically the same in most situations of interest.
Theoretical results call attention to the fact
that depending on the electron's
velocity, the background thermal-bath contribution 
can be  enhanced (or damped) by several
orders of magnitude \cite{CM1}.  This result was obtained in 
a simplified context by modeling the electron's spin flip 
by the transition of a two--level scalar system \cite{BL} 
coupled to the background thermal bath. The influence of the
velocity in the thermal depolarization rate 
can be understood by noticing that because of the Doppler 
effect the energy spectrum of the background photons is 
shifted in the electron's proper frame.
Thus, depending on the electron's velocity, photons 
of the background thermal bath can have their frequency shifted 
{\em into}  or {\em off} the absorbable band, implying  thus an
{\em enhancement} or {\em damping} of the excitation rate.
Although the two--level  model is a satisfactory approximation 
in many respects, this is incomplete in some other ones \cite{Betal2}. 
Here we aim to analyze the influence of the background thermal bath on 
true
fast-moving spin-1/2 fermions.

The paper is organized as follows:
In Section \ref{sec:angledistribution}, we calculate the angular 
distribution of emitted and absorbed photons, and radiated power 
induced by the spin flip. We show that in a certain interval of
solid angles the photon emission induced by the spin flip is
enhanced by several orders of magnitude because of the presence
of the thermal bath.
In Section \ref{sec:frequencydistribution} we exhibit the frequency 
distribution.
Section \ref{sec:totaldistribution} is devoted to calculate the 
total emission rate and
total radiated power induced by the spin flip. 
In Section \ref{sec:polarization} we
use previous section results to calculate the background
thermal bath influence on the depolarization of electrons at LEP. We show
that in spite of Sec.  \ref{sec:angledistribution} results,
the overall depolarization because of the background thermal
bath at LEP conditions is much smaller than the one because of
plain acceleration at zero-temperature.
Finally we discuss our results in Section \ref{sec:discussion}.
Natural units $\hbar = c= k =1$ will be adopted throughout the
paper.  

\setcounter{equation}{0}


\section{Photon angle distribution}
\label{sec:angledistribution}

In order to calculate the angle distribution of emitted photons 
induced by the spin flip of a fast-moving electron, 
it is useful to define from the beginning spherical angular coordinates
$(\theta, \phi)$ in an inertial frame at rest with the
laboratory, and with its origin instantaneously 
on the electron as follows: $\theta$ is the angle between the
electron's 3-velocity ${\bf v}$, 
and the 3-momentum ${\bf k}$ of the emitted photon,
while $\phi$ is the angle between the projection of ${\bf k}$ on
the plane orthogonal to ${\bf v}$,  and the electron's
3-acceleration ${\bf a}$. 

To calculate {\em at the tree level} the angular 
distribution of emitted and absorbed
photons associated with the spin-flip as well as the
corresponding radiated power, 
rather than using the thermal Green function approach,
we will introduce directly the proper 
thermal factor (Planck factor)  in the {\em vacuum} probability
distribution previously
calculated by Jackson \cite{J}: 
The photon emission rate {\em per laboratory time} 
$d{\cal P}^{\rm em}_{\rm vac}$ per solid angle 
$d\Omega =\sin \phi d\theta d\phi$, 
and frequency d$\omega$ 
induced by the spin flip of an electron circulating in a
storage ring at {\em zero-temperature} is
\begin{eqnarray}
\frac{ d^2 {\cal P}^{\rm em}_{\rm vac}(\theta_0)}{d\Omega d\omega} 
=  \frac{3 \sqrt{3}}{40 \pi^3} \frac{\nu^3 (1+t^2) }{\tau_0
\gamma^2 \omega_0} 
&\{& 
\sin^2 \theta_0 K^2_{1/3} (\eta )
+ \frac{1}{2} (1+\cos^2 \theta_0) (1+t^2) 
 [ K^2_{1/3} (\eta ) + K^2_{2/3} (\eta )]
\nonumber \\
 & + & 2 \cos \theta_0 \sqrt{1+t^2}
K_{1/3} (\eta ) K_{2/3} (\eta ) \},
\label{KEY}
\end{eqnarray}
where $\gamma = 1/\sqrt{1-v^2}$, $t = \gamma \theta \sin \phi$,
$\omega_0 $ is the electron's orbital frequency, 
\begin{equation}
\tau_0 = 
\left[ \frac{5\sqrt 3}{8} \frac{e^2 \gamma^5}{m^2 \rho^3}\right]^{-1}
\label{tau0}
\end{equation}
is the typical time interval for the electron beam to reach 
polarization equilibrium $P_0$, i.e. $P(t) =P_0 [1 -\exp(-t/\tau_0)]$, 
$m$ is the electron's mass,
$\rho$ is the bending radius of the storage ring, and     
$ \eta = \nu (1+t^2)^{3/2} /2 $ with
\begin{equation}
\nu \equiv \frac{2 \omega}{3 \gamma^3 \omega_0} .
\label{NI}
\end{equation}
For LEP we have $\gamma \approx 10^5$, $\omega_0 \approx 10^5 s^{-1}$
and a background temperature of $\beta^{-1} \approx 4 \cdot 10^{13} s^{-1}
(= 300 K)$.
The variable $\theta_0$ is the angle between the measurement direction of 
spin and  magnetic field before the transition. After any transition the
angle between spin and magnetic field changes to $\pi - \theta_0$.
Deexcitation  processes are characterized by the fact that
$ 0\le \theta_0 < \pi/2 $, while excitation  processes 
are characterized by the fact that $ \pi/2 < \theta_0 \le \pi $. 
We also recall that, at the tree level, each spin flip in
the vacuum is associated with a photon emission \cite{J}.

In the case the electron is moving in a
background thermal bath characterized by a temperature $\beta^{-1}$, 
the emission rate
can be expressed simply in terms of Eq. (\ref{KEY}) by (see e.g.
\cite{LT} for an account on photon radiation in a  heat bath)
\begin{equation}
\frac{ d^2 {\cal P}^{\rm em} (\theta_0)}{d\Omega d\omega}  =
\frac{ d^2 {\cal P}^{\rm em}_{\rm vac}(\theta_0)}{d\Omega d\omega} 
+\frac{ d^2 {\cal P}^{\rm em}_{\rm ther}(\theta_0)}{d\Omega d\omega}, 
\label{etfa}  
\end{equation}
where
$$
\frac{ d^2 {\cal P}^{\rm em}_{\rm ther}(\theta_0)}{d\Omega d\omega}
= 
\frac{ d^2 {\cal P}^{\rm em}_{\rm vac}(\theta_0)}{d\Omega d\omega}
n(\omega )
$$
with $n(\omega ) = 1/(e^{\omega \beta} -1)$  
accounts for stimulated emission.

In the presence of a background thermal bath, the spin-flip
process can be also related with the absorption of a photon. 
In order to calculate the absorption rate, we note that
because of unitarity, the absorption probability 
with spin-excitation (-deexcitation) 
must be equal to the {\em stimulated} emission probability 
with spin-deexcitation (-excitation):
\begin{equation}
\frac{ d^2 {\cal P}^{\rm abs} (\theta_0)}{d\Omega d\omega}  =
\frac{ d^2 {\cal P}^{\rm em}_{\rm ther}( \pi- \theta_0)}{d\Omega d\omega} 
.
\label{atfa}  
\end{equation}
The total spin-flip probability  will be given by summing up (\ref{etfa}) 
and
(\ref{atfa}), and integrating the result over the frequency 
$\omega$ and solid angle  $\Omega$ as shown in Sec. 
\ref{sec:totaldistribution}.

\begin{figure}[htb]
\epsfxsize=0.9\textwidth
\begin{center}
\leavevmode
\epsfbox{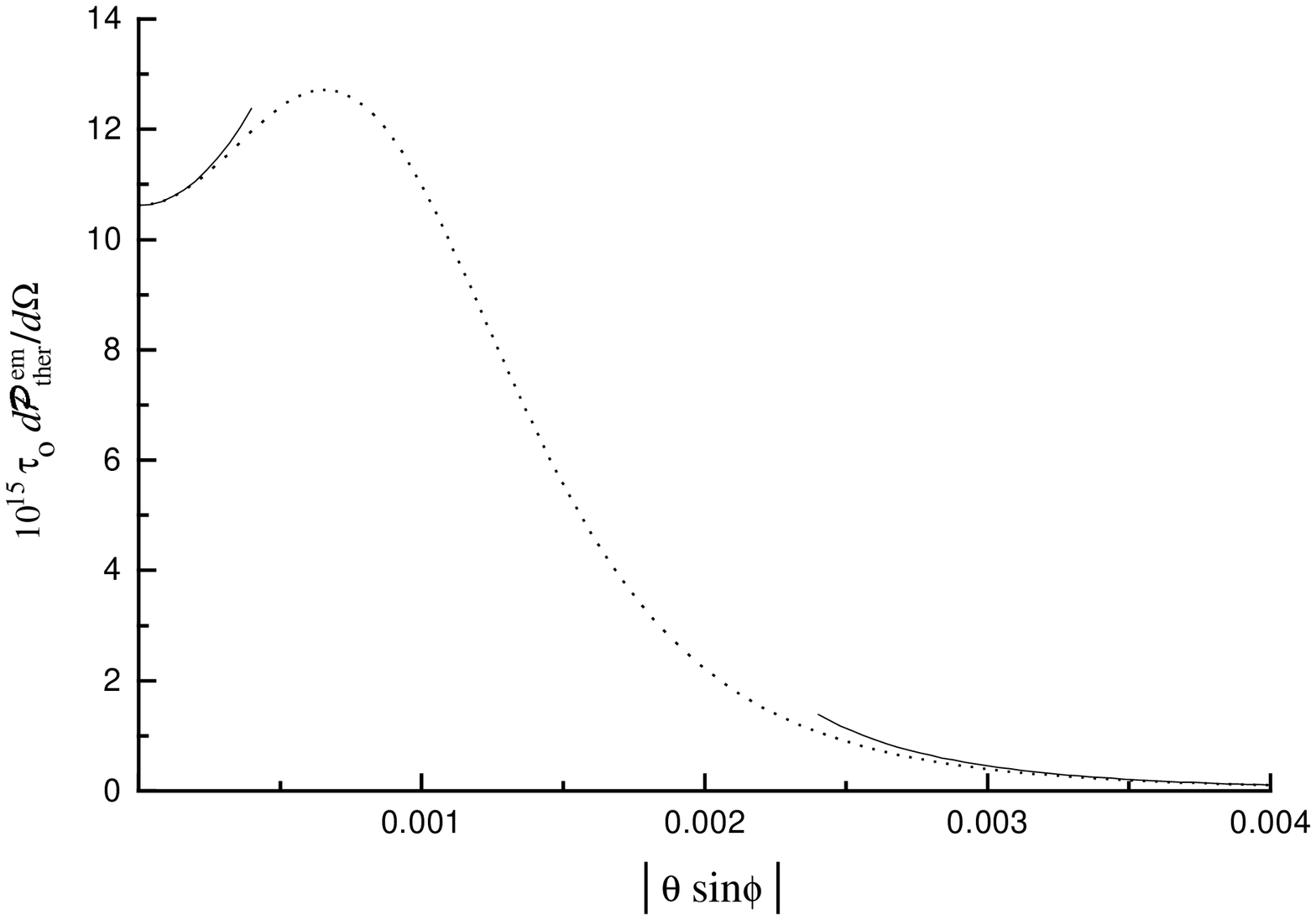}
\end{center}
\caption{Thermal contribution to the 
angular distribution of radiation induced 
by the deexcitation of electrons at LEP. The dashed line was obtained
through numerical integration and is to be compared with full line
obtained through analitic approximation. The analogous figure for
excitation is very similar.}
\label{EMRATEPERANGLE}
\end{figure}

\begin{figure}[htb]
\epsfxsize=0.9\textwidth
\begin{center}
\leavevmode
\epsfbox{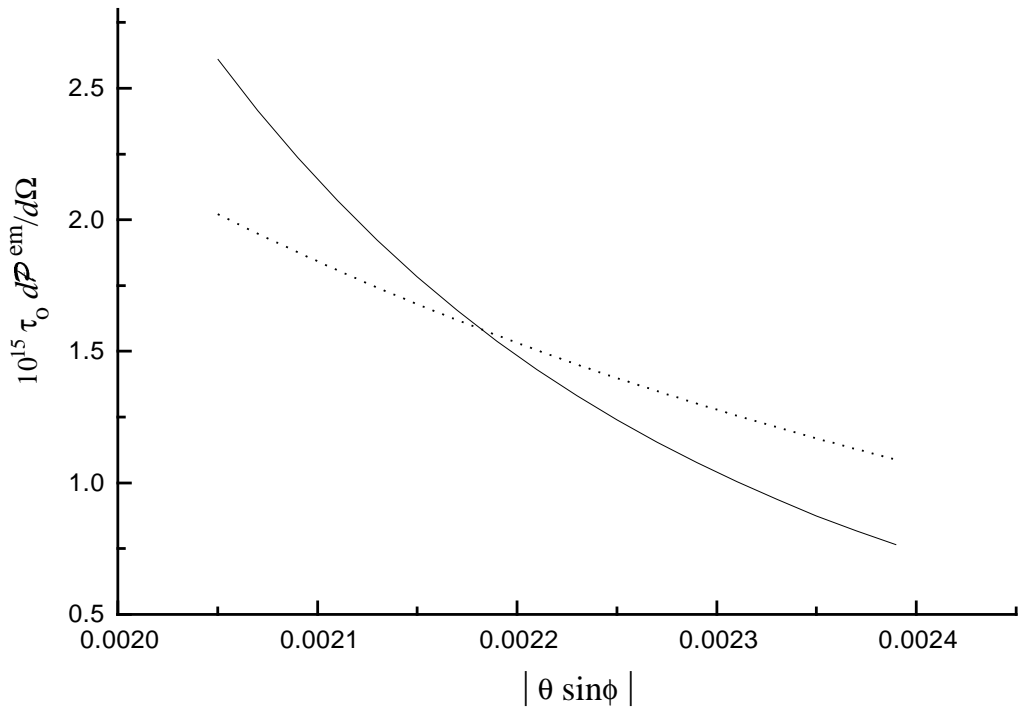}
\end{center}
\caption{The dashed line represents 
{\protect{$ d {\cal P}^{\rm em}_{\rm ther} (\theta_0)/ d\Omega  $ }}
while the full line represents
{\protect{$ d {\cal P}^{\rm em}_{\rm vac} (\theta_0)/ d\Omega  $}}.
For sufficiently ``large'' $|\theta \sin \phi |$, the spin-flip photon
emission is  dominated by the presence of the thermal bath.}
\label{PRINCIPAL}
\end{figure}

In order to obtain the angle distribution of emitted 
photons, we  integrate (\ref{etfa}) over frequencies $\omega$.  
In the $|t|<50$ interval, we use the approximation 
$K_{a > 0} (\eta \ll 1) \approx {\Gamma(a ) 2^{a -1}}/{\eta^a}$
since $n (\omega \beta \gg 1 ) \sim  e^{-\omega \beta}$  implies
that the integral only collaborates significantly 
for  
$0<\omega\beta < 10$
and in this range $\eta \ll 1$.
In the $|t| >9 \cdot 10^2$ interval, we use the approximation
$n(\omega \ll 1/\beta) \approx 1/\beta \omega$ 
since
$K_{a \ge 0} (\eta \gg 1) \sim \sqrt{\pi/2\eta } \; e^{-\eta}$
implies that the
integral only collaborates significantly for 
$0<\eta < 10$, and in this range
$  \omega \ll 1/\beta $. Hence,
after some algebra we obtain 
\begin{equation}
\frac{  d  {\cal P}^{\rm em} (\theta_0 )}{d\Omega } 
 = 
\frac{ d {\cal P}^{\rm em}_{\rm vac}(\theta_0)}{d\Omega }
+  
\frac{ d {\cal P}^{\rm em}_{\rm ther}(\theta_0)}{d\Omega }
\label{R1}
\end{equation}
where
\begin{eqnarray}
\left. 
\frac{d{\cal P}^{\rm em}_{\rm ther}(\theta_0 )}{d\Omega } 
\right|_{|t|<50} 
&  & = 
\frac{\Gamma^2 (2/3) \Gamma (8/3) \zeta (8/3)}{5 \cdot 48^{1/6} \pi^3 }
\frac{\gamma^{-7}}{\tau_0 } 
\left(\frac{\beta^{-1}}{\omega_0 }\right)^{8/3}
\left\{ 
\frac{ \Gamma^2 (1/3) \Gamma (10/3) \zeta(10/3)}{6^{2/3} \Gamma^2 (2/3)
\Gamma(8/3) \zeta(8/3) }
\right.  
\nonumber \\
& \times &
\frac{ \sin^2 \theta_0 }{ \gamma^2}
\left( \frac{\beta^{-1}}{ \omega_0} \right)^{2/3} 
+ \left( \frac{32}{3} \right)^{1/3}
\frac{\Gamma(1/3)  \zeta(3)}{\Gamma (2/3) \Gamma (8/3) \zeta(8/3)}
\frac{\cos \theta_0 }{\gamma }
\left( \frac{\beta^{-1}}{ \omega_0 }\right)^{1/3}
\nonumber \\
& + &
\left.
\frac{(1+ \cos^2 \theta_0)}{2} 
\left[
\frac{ \Gamma^2 (1/3) \Gamma (10/3) \zeta (10/3)}{ 6^{2/3} \Gamma^2 (2/3) 
\Gamma (8/3) \zeta (8/3)  } 
\frac{(1+t^2)}{\gamma^2}
\left( \frac{\beta^{-1}}{\omega_0 }\right)^{2/3} +1
\right]
\right\} 
\label{fim1}   
\end{eqnarray}
\begin{eqnarray}
\left. 
\frac{ d {\cal P}^{\rm em}_{\rm ther}(\theta_0)}{d\Omega } 
\right|_{|t| >9 \cdot 10^2} 
&=& 
\frac{\sqrt 3 }{48 \pi}  
\frac{(1+ t^2)^{-7/2}}{\tau_0 \omega_0 \beta \gamma^2} 
\nonumber \\
&\times&  
\left\{ 
\sin^2 \theta_0 +\frac{6}{5} (1+ \cos^2
\theta_0)( 1 +t^2) +\frac{64}{5 \sqrt 3 \pi}  \sqrt{1+t^2} \cos
\theta_0 
\right\} ,
\label{fim2}  
\end{eqnarray}
and 
\begin{equation}
\frac{ d {\cal P}^{\rm em}_{\rm vac}(\theta_0)}{d\Omega } 
=
 \frac{16 }{45 \pi^2 }  
 \frac{\gamma (1+ t^2)^{-5}}{\tau_0 }  
\left\{
\sin^2 \theta_0 +\frac{9}{8} (1+ \cos^2
\theta_0)( 1 +t^2) +\frac{105 \sqrt 3 \pi}{256}  \sqrt{1+t^2}
\cos \theta_0 
\right\}.
\label{J65}  
\end{equation}
Eqs. (\ref{fim1}) and (\ref{fim2}) are plotted in Fig.
\ref{EMRATEPERANGLE} over
the result obtained through explicit numerical integration, and
are in perfect agreement. Fig. \ref{PRINCIPAL} 
plots 
$  
{ d {\cal P}^{\rm em}_{\rm ther}(\theta_0)}/{d\Omega }
$
against
$
{ d {\cal P}^{\rm em}_{\rm vac}(\theta_0)}/{d\Omega } ,
$
and shows that for ``large'' $\theta \sin \phi$, the spin-flip photon
emission is largely dominated by the presence of the thermal bath.  
In particular at LEP for $|t| \approx 10^5$
($\theta = \phi = \pi/2$), we have
$  
({ d {\cal P}^{\rm em}_{\rm ther}(\theta_0)}/{d\Omega })/ 
({ d {\cal P}^{\rm em}_{\rm vac}(\theta_0)}/{d\Omega }) \approx 10^8.
$
{\em This shows that the background
thermal bath must not be always overlooked here.}

\begin{figure}[htb]
\epsfxsize=0.9\textwidth
\begin{center}
\leavevmode
\epsfbox{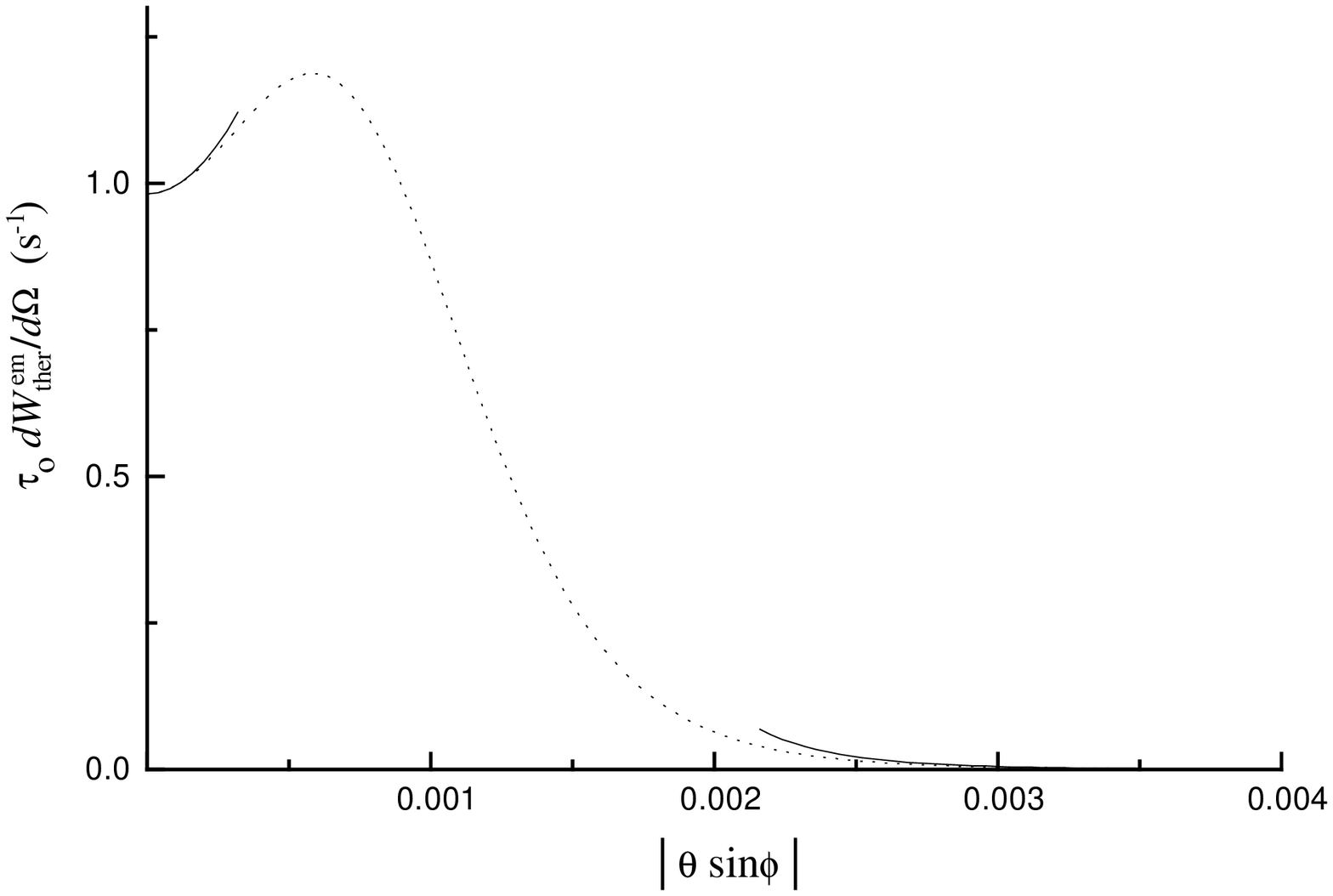}
\end{center}
\caption{Thermal contribution to the 
angular-distribution of the radiated power induced 
by the deexcitation of electrons at LEP. The dashed line was obtained
through numerical integration and is to be compared with full curves
obtained through analitic approximations. The analogous figure for 
excitation is very similar.}
\label{POWERPERANGLE1}
\end{figure}

\begin{figure}[htb]
\epsfxsize=0.9\textwidth
\begin{center}
\leavevmode
\epsfbox{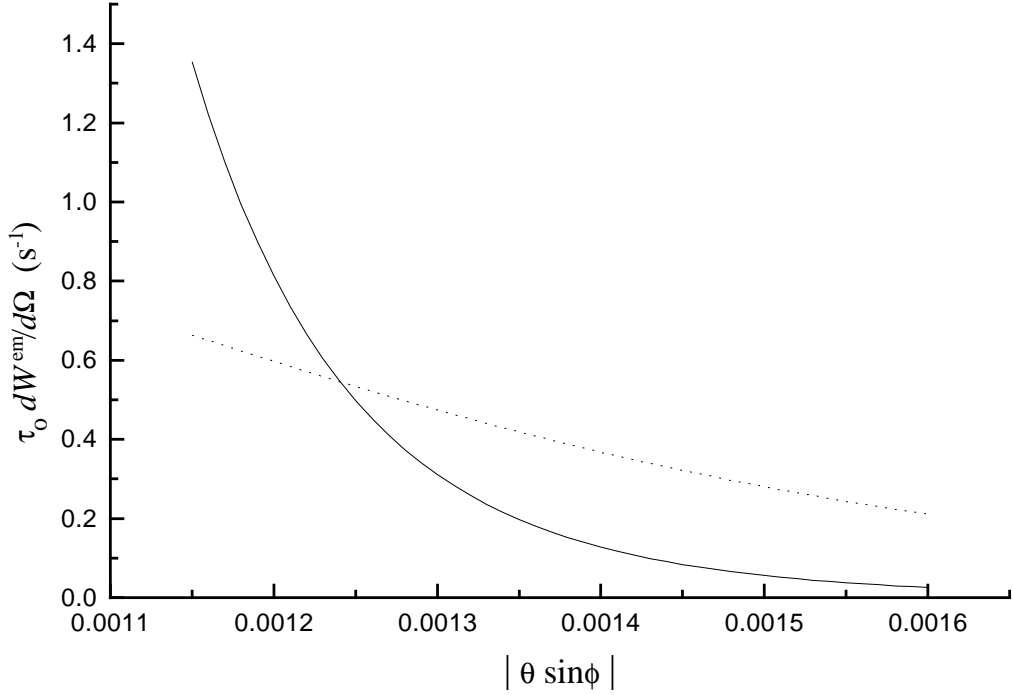}
\end{center}
\caption{The dashed line represents 
\protect{$  { d {\cal W}^{\rm em}_{\rm ther}(\theta_0)}/{d\Omega } $},
while the full line represents
\protect{${ d {\cal W}^{\rm em}_{\rm vac}(\theta_0)}/{d\Omega }$.}
For sufficiently ``large'' $|\theta \sin \phi |$, the radiated power 
is dominated by the presence of the thermal bath.}
\label{POWERPERANGLE2}
\end{figure}

The angular distribution of the
radiated power is obtained by multiplying (\ref{etfa}) by $\omega$ and
integrating over frequencies. By using the same approximations
described above, we obtain
\begin{equation}
\frac{  d  {\cal W}^{\rm em}(\theta_0 )}{d\Omega } 
= 
\frac{ d {\cal W}^{\rm em}_{\rm vac}(\theta_0)}{d\Omega }
+  
\frac{ d {\cal W}^{\rm em}_{\rm ther}(\theta_0)}{d\Omega } ,
\label{R2}
\end{equation}
where (see Figs. \ref{POWERPERANGLE1}-\ref{POWERPERANGLE2})
\begin{eqnarray}
\left.
\frac{  d  {\cal W}^{\rm em}_{\rm ther}(\theta_0 )}{d\Omega } 
\right|_{|t| <50}
& = &
\frac{\sqrt 3}{45 \pi^3 }
\frac{\omega_0}{\tau_0 \gamma^7 }
\left(\frac{\beta^{-1}}{ \omega_0} \right)^{11/3}
\left\{ 
\left(\frac{3}{4}\right)^{2/3}
\Gamma^2(1/3) \Gamma (13/3)  \zeta(13/3)
\frac{ \sin^2\theta_0 }{ \gamma^2} 
\left( \frac{\beta^{-1}}{ \omega_0} \right)^{2/3}
\right.  
\nonumber \\
& + &
\left(\frac{9}{32}\right)^{1/3}
(1+ \cos^2 \theta_0)
\left[
\frac{\Gamma^2 (1/3)\Gamma (13/3) \zeta (13/3)}{ 2^{2/3} } 
\frac{(1+t^2)}{ \gamma^2}
\left( \frac{\beta^{-1}}{  \omega_0 } \right)^{2/3}
\right.
\nonumber \\
& + &
\left.
\left.
\frac{ \Gamma^2 (2/3) \Gamma (11/3) \zeta (11/3)}{3^{-2/3}} 
\right]
+ \frac{36 \pi \zeta(4)}{\sqrt 3 } 
\frac{\cos \theta_0}{\gamma }
\left( \frac{\beta^{-1}}{ \omega_0 } \right)^{1/3}
\right\},
\label{fim3}  
\end{eqnarray}
\begin{eqnarray}
\left. 
\frac{ d {\cal W}^{\rm em}_{\rm ther}(\theta_0)}{d\Omega } 
\right|_{|t| >9 \cdot  10^2} 
&=& 
 \frac{16 }{45 \pi^2 }  
 \frac{\gamma (1+ t^2)^{-5}}{\tau_0 \beta}  
 \left\{
\sin^2 \theta_0 +\frac{9}{8} (1+ \cos^2
\theta_0)( 1 +t^2)
 \right.
\nonumber \\
& + &
\left.  
\frac{105 \sqrt 3 \pi}{256}  \sqrt{1+t^2}
\cos \theta_0 
\right\} ,
\label{fim4}  
\end{eqnarray}
and
\begin{equation}
\frac{ d {\cal W}^{\rm em}_{\rm vac}(\theta_0)}{d\Omega } 
=
 \frac{77 \sqrt 3 }{256 \pi }  
 \frac{\gamma^4 \omega_0 }{(1+ t^2)^{13/2} \tau_0 }  
\left\{
\sin^2 \theta_0 +\frac{12}{11} \frac{(1+ \cos^2
\theta_0)}{ (1 +t^2)^{-1}} +\frac{8192 \sqrt 3 }{2079 \pi}  \sqrt{1+t^2}
\cos \theta_0 
\right\} .
\label{J66}  
\end{equation}

\setcounter{equation}{0}


\section{Frequency distribution}
\label{sec:frequencydistribution}

The frequency distribution of emitted 
photons can be obtained by integrating (\ref{etfa}) in the solid
angle. By using the approximation \cite{J} $ d\Omega \approx
(2\pi /\gamma) dt$ which is good for small $\theta$, we obtain:
\begin{eqnarray}
\frac{ d {\cal P}^{\rm em} (\theta_0)}{d\omega } 
&=&
 \frac{3}{10 \pi } \frac{\nu^2}{\gamma^3 \omega_0 \tau_0}  
\left[ 
 \sin^2 \theta_0 \int_\nu^{\infty} K_{1/3} (s) ds
+   ( 1 + \cos^2 \theta_0) K_{2/3} (\nu ) 
\right.
\nonumber \\
&+& 
\left.
2 \cos \theta_0 K_{1/3} (\nu)
\right] \; [1+ n(\omega )].
\label{fim5}  
\end{eqnarray}
The small-angle approximation above is corroborated by the last
section results (see Figs. \ref{EMRATEPERANGLE}-\ref{PRINCIPAL}).
The unit in the square brackets is related with the vacuum (see Ref.
\cite{J}) and accounts for {\em spontaneous} emission, while the
$n(\omega )$ term is related with the background thermal bath and
accounts for {\em stimulated} emission.

The frequency distribution of the
radiated power is trivially obtained from this result by simply
multiplying (\ref{fim5}) by $\omega$, and is
introduced for sake of completeness:
\begin{eqnarray}
\frac{ d {\cal W}^{\rm em} (\theta_0)}{d\omega } 
&=&
 \frac{3}{10 \pi } \frac{\nu^2 \omega}{\gamma^3 \omega_0 \tau_0}  
\left[ 
 \sin^2 \theta_0 \int_\nu^{\infty} K_{1/3} (s) ds
+   ( 1 + \cos^2 \theta_0) K_{2/3} (\nu ) 
\right.
\nonumber \\
&+& 
\left.
2 \cos \theta_0 K_{1/3} (\nu)
\right] \; [1+ n(\omega )].
\label{fim6}  
\end{eqnarray}
As a lateral comment, we note that for 
$\omega \beta < \ln 2 $ the background thermal contribution dominates 
over the
vacuun term.
These results will be used in the next section to calculate
the total photon emission, and power radiated.
\setcounter{equation}{0}


\section{Total emission rate and radiated power}
\label{sec:totaldistribution}

\setcounter{equation}{0}

In order to calculate the total photon emission rate  and
radiated power, we integrate (\ref{fim5}) and (\ref{fim6}) in
frequencies. The vacuum term is trivially integrated.
For LEP parameters and 
$\gamma > 3 \cdot 10^3 $, in order to
integrate the thermal term, we use the approximation
$K_{a >0 } (\nu << 1) \approx {\Gamma(a ) 2^{a -1}}/{\nu^a}$,
since $n(\omega \beta \gg 1) \sim e^{- \omega \beta} $
implies that the integral only collaborates significantly 
for 
$0< \omega \beta < 10 $ and in this interval $\nu \ll 1$. 
Now, if 
$10$ 
{\raise0.3ex\hbox{$\;<$\kern-0.75em\raise-1.1ex\hbox{$\sim\;$}}}
$\gamma $
{\raise0.3ex\hbox{$\;<$\kern-0.75em\raise-1.1ex\hbox{$\sim\;$}}}
$10^2$, 
in order to integrate the thermal term we use the
approximation $n(\omega \ll 1/\beta ) \approx 1/(\omega \beta)$
since  
$K_{a \ge 0} (\nu \gg 1) \sim \sqrt{\pi/2\nu } \; e^{-\nu}$
implies that the integral only collaborates significantly 
for 
$0<\nu <10$, and
in this interval $\omega \ll 1/\beta$. In doing these
approximations, one must keep in mind that (\ref{KEY}) and our
last section's assumption $d\Omega \approx (2\pi /\gamma ) dt $  
are only valid in relativistic regimes.
In summary, we obtain for the total emission rate
\begin{equation}
{\cal P}^{\rm em}(\theta_0 ) 
= 
{\cal P}^{\rm em}_{\rm vac}(\theta_0)
+  
{\cal P}^{\rm em}_{\rm ther}(\theta_0) ,
\label{R3}
\end{equation}
where
\begin{equation}
{\cal P}^{\rm em}_{\rm ther}(\theta_0)
=
\frac{1}{2\tau_0}\left( 4\cdot 10^{19} \gamma^{-7} + 6 \cdot
10^{22} \gamma^{-8} 
\cos \theta_0 \right)
\label{R5}
\end{equation}
for $\gamma> 3 \cdot 10^3$;
\begin{equation}
{\cal P}^{\rm em}_{\rm ther}(\theta_0)
=
\frac{1}{2\tau_0} \frac{8 \cdot 10^8}{5 \gamma^3} 
\left(  \frac{2 }{\sqrt 3} + \cos \theta_0 \right)
\label{R6}
\end{equation}
for 
$10$ 
{\raise0.3ex\hbox{$\;<$\kern-0.75em\raise-1.1ex\hbox{$\sim\;$}}}
$\gamma $
{\raise0.3ex\hbox{$\;<$\kern-0.75em\raise-1.1ex\hbox{$\sim\;$}}}
$10^2$; and
\begin{equation}
{\cal P}^{\rm em}_{\rm vac}(\theta_0)
=
\frac{1}{2\tau_0}\left( 1 + \frac{8}{5 \sqrt 3} \cos \theta_0
\right) 
\label{R4}
\end{equation}
for any $\gamma$,
where we assume $\theta_0 = 0$ for deexcitation and $\theta_0
=\pi$ for excitation because hereafter we will suppose 
the polarization to be measured along the magnetic field direction. 
Analogously, we obtain for the total radiated power
\begin{equation}
{\cal W}^{\rm em}(\theta_0 ) 
= 
{\cal W}^{\rm em}_{\rm vac}(\theta_0)
+  
{\cal W}^{\rm em}_{\rm ther}(\theta_0) ,
\label{W3}
\end{equation}
where
\begin{equation}
{\cal W}^{\rm em}_{\rm ther}(\theta_0)
=
\frac{1}{2\tau_0}\left(4 \cdot 10^{33} \gamma^{-7} + 5 \cdot
10^{36} \gamma^{-8}  
\cos \theta_0 \right)
\label{W5}
\end{equation}
for $\gamma> 3 \cdot 10^3$ ;
\begin{equation}
{\cal W}^{\rm em}_{\rm ther}(\theta_0)
=
\frac{4 \cdot 10^{13}}{2\tau_0}   
\left( 1  + \cos \theta_0 \right)
\label{W6}
\end{equation}
for $10 < \gamma <10^2$; and
\begin{equation}
{\cal W}^{\rm em}_{\rm vac}(\theta_0)
=
\frac{4 \cdot 10^{5} \gamma^3}{2\tau_0}\left( 1 +  \cos \theta_0
\right) 
\label{W4}
\end{equation}
for any $\gamma$.
In particular, for $\gamma =10^5$ (LEP) we have
\begin{equation}
{\cal P}^{\rm em}_{\rm ther}(\theta_0)
=
\left( 4\cdot 10^{-16} + 6 \cdot 10^{-18} 
\cos \theta_0 \right) / {2\tau_0}, \;\;\;\;\;
{\cal W}^{\rm em}_{\rm ther}(\theta_0)
=
\left( 4 \cdot  10^{-2} + 5 \cdot 10^{-4} 
\cos \theta_0 \right) / {2\tau_0}
\label{+}
\end{equation}
which are much smaller than 
\begin{equation}
{\cal P}^{\rm em}_{\rm vac}(\theta_0)
=
\left( 1 + 9 \cdot 10^{-1} 
\cos \theta_0 \right) / {2\tau_0}, \;\;\;\;\;
{\cal W}^{\rm em}_{\rm vac}(\theta_0)
=
\left( 4 \cdot  10^{20} + 4 \cdot 10^{20} 
\cos \theta_0 \right) / {2\tau_0}
\label{++}
\end{equation}
respectively.  This result shows that
eventually the background thermal-bath contribution to the total
transition rate is very small in this case, and can be disregarded for
depolarization purposes. This will be explicitly shown in
the next section. Note, however, the strong $\gamma$
dependence on ${\cal P}^{\rm em}_{\rm ther}(\theta_0)$ and
${\cal W}^{\rm em}_{\rm ther}(\theta_0)$ which 
makes the thermal contribution larger than the vacuum
contribution in the $10 <\gamma < 10^2$ range. As a consequence,
the background thermal bath not only is important to the
photon-emission rate for large solid angles at LEP-kind accelerators
as shown in Sec. \ref{sec:angledistribution},
but could be also important for the polarization itself provided
$\gamma$ was considerably smaller. 


\section{Polarization}
\label{sec:polarization}

\setcounter{equation}{0}

Finally, let us calculate the polarization function 
\begin{equation}
P= \frac{{\cal P}_{\downarrow} - {\cal P}_{\uparrow}}{
{\cal P}_{\downarrow} + {\cal P}_{\uparrow} } ,
\label{PolFim}
\end{equation}
for electrons at LEP taking into account the background thermal
bath, where the excitation rate is given by
\begin{equation}
{\cal P}_\uparrow =  {\cal P}^{\rm em}_{\rm vac} (\theta_0 =\pi ) 
                    + {\cal P}^{\rm em}_{\rm ther} (\theta_0 =\pi )
                    + {\cal P}^{\rm abs}  (\theta_0 =\pi ) ,   
\label{PTOT}
\end{equation}
and the deexcitation rate is given by
\begin{equation}
{\cal P}_\downarrow =  {\cal P}^{\rm em}_{\rm vac} (\theta_0 =0 ) 
                    + {\cal P}^{\rm em}_{\rm ther} (\theta_0 =0 )
		    + {\cal P}^{\rm abs}  (\theta_0 =0 ) .   
\label{PDTOT}
\end{equation}
${\cal P}^{\rm em}_{\rm vac} (\theta_0 =\pi )$, 
${\cal P}^{\rm em}_{\rm ther} (\theta_0 =\pi )$
and
${\cal P}^{\rm abs}  (\theta_0 =\pi ) $
are the excitation rates associated with spontaneous photon emission,
stimulated photon emission and photon absorption respectively,
while
${\cal P}^{\rm em}_{\rm vac} (\theta_0 =0 )$, 
${\cal P}^{\rm em}_{\rm ther} (\theta_0 =0 )$
and      
${\cal P}^{\rm abs}  (\theta_0 =0 ) $
are the deexcitation rates associated analogously with spontaneous 
photon emission, stimulated photon emission and photon absorption.
 
Now, by substituting (\ref{PTOT}) and (\ref{PDTOT}) in
(\ref{PolFim}), and using (\ref{atfa}), we obtain
\begin{equation}
P \approx P_{\rm vac} 
\left( 
1 - 2\frac{
{\cal P}^{\rm em}_{\rm ther} (\theta_0 = 0) + {\cal P}^{\rm em}_{\rm 
ther} 
(\theta_0 = \pi)}{{\cal P}^{\rm em}_{\rm vac}
(\theta_0=0) + {\cal P}^{\rm em}_{\rm vac} (\theta_0 = \pi)}
\right) .
\end{equation}
where $P_{\rm vac} =0.92 $ is the vacuum polarization obtained
at zero temperature.
Finally, by using  (\ref{+}) and (\ref{++}), we obtain 
$$
P \approx P_{\rm vac} (1-8 \cdot 10^{-16}),
$$
which confirms last section's ``conjecture'' that the background thermal 
bath
contribution to the depolarization should be small. 


\section{Discussion}
\label{sec:discussion}

We have discussed the influence of the background thermal bath
on the depolarization of electrons in high-energy storage rings,
and the corresponding photon emission and radiated power. We
have calculated the angle and frequency distribution of such
photons and obtained  that in a large 
interval of solid angles the
photon emission is enhanced by several orders of magnitude
because of the thermal bath. In addition, we have shown that the 
background
thermal bath can be very important  
to the total photon emission and 
overall depolarization in some $\gamma$-interval, 
although it can be neglected at LEP and
similar accelerators. In spite of the fact that 
some of these conclusions were 
anticipated before \cite{CM1} by modeling the electron's spin flip by the
transition of a two-level scalar system, this approximate
approach and the exact calculation (at the tree level) here developed
lead to fairly different numerical results. This is another
indication of the outstanding role played by  Thomas precession
in this context as firstly called attention by Bell and Leinaas 
\cite{BL}, 
and further investigated in more detail by Barber et al \cite{Betal2}.

\acknowledgments 
We are really grateful to Desmond Barber for reading carefully
our manuscript, and for his enlightening comments.
AG and DV acknowledge full support by Funda\c c\~ao de Amparo
\`a Pesquisa do Estado de S\~ao Paulo, while GM was partially 
supported by Conselho
Nacional de Desenvolvimento Cient\'{\i}fico e Tecnol\'ogico.

\newpage


\end{document}